**Heating reduction as collective action: Impact on attitudes, behavior and energy consumption in a Polish field experiment**

Mona Bielig1,2, Lukasz Malewski3, Karol Bandurski3, Florian Kutzner1, Melanie Vogel1, Sonja Klingert4, Radoslaw Gorzenski3 & Celina Kacperski1,2

[1]Seeburg Castle University

[2]Centre for the Advanced Study of Collective Behavior, University of Konstanz

[3]Poznan University of Technology

[4]University of Stuttgart

Corresponding author: Mona Bielig, mona.bielig@uni-konstanz.de

Author Note

Financial support by the European Union Horizon 2020 research and innovation program is gratefully acknowledged (Project RENergetic, grant N957845).

Abstract

Heating and hot water usage account for nearly 80% of household energy consumption in the European Union. To reach the EU Green Deal goals, new policies to reduce heat energy consumption are indispensable. However, research on reducing energy consumption concentrates either on technical building interventions without considerations of people's behavior, or psychological interventions with no technical interference. Such interventions can be promising, but their true potential for scaling up can only be realized by testing approaches that integrate behavioral and technical solutions in tandem rather than in isolation. In this research, we study a mix of psychological and technical interventions targeting heating and hot water demand among students in Polish university dormitories. We evaluate effects on building energy consumption, behavioral spillovers and on social beliefs and attitudes in a pre-post quasi-experimental field study in three student dormitories. Our findings reveal that the most effective approaches to yield energy savings were a direct, collectively framed request to students to reduce thermostat settings for the environment, and an automated technical adjustment of the heating curve temperature. Conversely, interventions targeting domestic hot water had unintended negative consequences, including increased energy use and negative spillovers, such as higher water consumption. Further, we find that informing students about their active, collective participation had a positive impact on perceived social norms. Our findings highlight the importance of trialing interventions in controlled real-world settings to understand the interplay between technical systems, behaviors, and social impacts to enable scalable, evidence-based policies driving an effective and sustainable energy transition.

## 1. Introduction

The energy sector plays a significant role in climate change, contributing approximately three-quarters of global greenhouse gas emissions (IEA, 2021). Reducing these emissions requires both a phase-out of fossil fuels, but also a significant reduction in energy consumption. Heating and hot water usage account for almost 80% of total energy use in European Union households[1], making these a significant lever of potential reductions in energy consumption.

While building infrastructure and heating technology account for much variation in energy consumption, decisions and behaviors of building occupants can also achieve savings in energy (Bandurski et al., 2017; Janda, 2011). Exemplary, in single family houses, occupants and building characteristics account for a similar impact in heating consumption (Van Den Brom et al., 2019). More importantly, residents' consent is required for technical interventions that target heating systems to make use of the interventions' potential: For instance, research indicates lowering of a thermostat by just one degree Celsius can reduce energy consumption by approximately 7%[2].

This means that reducing global energy use requires, to a considerable degree, encouraging individuals to adopt more sustainable practices in their heating habits (Iweka et al., 2019; Lopes et al., 2012). Intervention targets can include not only daily decisions about interactions with energy-consuming devices and building systems, but also long-term decisions related to energy efficiency investments (Heydarian et al., 2020; Matsumoto & Sugeta, 2022). Effective household saving strategies therefore demand a policy focus on behavioral science insights to aid local, experimental collaborations integrating efforts across academia, industry, and government (Berger et al., 2023).

[1] https://energy.ec.europa.eu/topics/energy-efficiency/energy-efficient-buildings/energy-performance-buildings-directive_en
[2] https://energy.ec.europa.eu/topics/markets-and-consumers/actions-and-measures-energy-prices/playing-my-part_en

While interventions for behavior changes in the electricity domain have been widely studied (Composto & Weber, 2022), the potential for reducing energy consumption through behavioral interventions in heating and hot water systems is less explored. This is especially important since there is some indication that heating-related behaviors are less amenable to change than those related to electricity use (Spence et al., 2015) and require more commitment due to reductions in comfort (Barr et al., 2005). In our research we therefore focus on the heating and hot water usage behavior of students residing in dormitories at a Polish university, to test the effectiveness of interventions targeting both the active behavior, and the technical system underlying their heating and hot water use.

## 2. Conceptual background

Behavior change can be achieved through targeted interventions, which are systematic activities aimed at modifying behavior patterns. These activities can include providing incentives, offering support, disseminating information, and giving guidance (Michie et al., 2011) or directly influencing the systems within which occupants interact. Methodologically, we consider two ways of interfering with energy consumption of heating and hot water: one, with behavioral interventions targeting individuals themselves, e.g. by motivating them to reduce their thermostat setpoint, or two, with technical interventions targeting the pre-defined parameters of the installed system, for which in many cases still acceptance of occupants is still helpful or even needed. In our field trial, we test both intervention types relying on an existing Building Management System in three student dormitories. We collect pre-post measurements and assign different dormitories to different intervention groups, assessing the effects of these interventions on energy and water consumption, as well as social impact in the form of students' attitudes, with a within-between quasi-experimental design. By testing both technical and behavioral interventions in an interdisciplinary approach, we aim not only to evaluate their effectiveness but also provide critical insights for scaling these strategies to broader contexts (Berger et al., 2023).

### 2.1. Changing heating and hot water usage in the field – behavioral science perspective

A scoping review on the effectiveness of behavioral interventions to reduce household energy consumption shows that interventions and their effects vary strongly between behavioral tools used and behaviors targeted – with most studies focusing on reduction of electricity consumption (Composto & Weber, 2022). Evidence from trials like ours, with a focus on reduction in heating and hot water usage in a field setting, is scarce but seems promising. In one field intervention, using a thermal image for energy awareness and visualization was successful in reducing household heating demand over a year, leading to a significant reduction in $CO_2$ (Goodhew et al., 2015). A different field study demonstrated that a motivational intervention, which involved reflecting on personal reasons to save energy and setting energy goals, led to a substantial decrease in hot water use (Legault et al., 2020). An intervention combining real-time feedback with social comparison in a Swiss energy-efficient district successfully led to a reduction in hot water consumption (Tomic et al., 2024) and hot water usage in hand washing was reduced using vivid messages in an immersive virtual environment (Bailey et al., 2015). In a long-term field study in Belgian cities, an informational behavior change intervention to reduce energy led to a significant decrease in gas consumption (Lange et al., 2024). Notably, all these interventions did not require a financial incentive to change behavior, but motivated individuals with symbolic incentives, which have been shown to be very effective in the electricity context as well (Kacperski et al., 2022; Kacperski & Kutzner, 2020).

As a consequence of behavioral interventions the targeted energy behavior itself might change, but also other subsequent energy behaviors or broader pro-environmental behaviors (Nilsson et al., 2017): the initial behavior change influencing the probability for a subsequent one is known as a spillover effect, with desirable changes termed a positive, undesirable ones a negative spillover (Nilsson et al., 2017; Truelove et al., 2014). Engaging in an initial pro-environmental action can promote and motivate further actions (positive behavioral spillover), as demonstrated in a study where a water saving campaign did not only reduce water usage but also decreased electricity consumption (Carlsson et al., 2021). Following the review of Truelove et al. (2014), positive spillovers are particularly likely when the action informs one's identity. On the other hand, the engagement in one pro-environmental action can also reduce the likelihood of a following behavior, a negative spillover: in contrast to the prior example, a different study found that a water saving intervention led to a reduction in water use, but caused an increase in energy usage (Tiefenbeck et al., 2013). In our study, we will examine spillover effects to other energy domains, i.e. different behavioral energy consumption measures when implementing interventions targeting space heating and hot water. While these interventions could either (a) be accepted and thus reach the targeted goal of reducing energy consumption or even strengthen pro-environmental behaviors in other energy domains, they could also lead to compensatory spillover behaviors through (b) increase of space heating or water temperature, (c) using more water or (d) even compensating through other devices like electrical heating.

### 2.2. Social impact

A pro-environmental intervention might lead to changes in perceived social identity processes and thereby impact further pro-environmental intentions: for example, participating in a water saving intervention changed students' efficacy beliefs and intentions for different pro-environmental behaviors in a field study in the UK (Haggar et al., 2023). A review has shown that social impacts, though considered highly important across many domains, are under-researched in the literature on the impact of energy collective actions (Bielig et al., 2022). To assess the social impact of our interventions, going beyond solely assessing the energy impact and behavioral spillovers, we conducted a survey before and after the introduction of the space heating interventions. This allows us to gain a more holistic picture of intervention impact, considering also the effects on attitudes and beliefs through participating in the intervention. We assessed not only the attitude towards the intervention itself, but also included measures building on established models for pro-environmental actions, such as the SIMPEA (Fritsche et al., 2018), namely social norms, collective efficacy beliefs and social identification. We framed participation in our interventions as a collective action, targeting the pre-existing local social identity of being a student of the university. Former research shows that such local, social identities can influence participation in pro-environmental, and collective actions (Bielig et al., 2024; Fritsche et al., 2018; Kacperski et al., 2023; van Zomeren et al., 2018). Additionally, we assessed compensatory green beliefs (Penker & Seebauer, 2023) to better understand potential spillover effects, and pro-environmental policy support as a measure for intentional spillover.

### 2.3. Changing space heating and domestic hot water usage in the field – a building perspective

Next to behavioral science informed interventions, changing the energy consumed by space heating and domestic hot water can also be achieved by directly targeting the underlying technical systems, through efficiency or conservation interventions in buildings. In Building Management Systems, space heating (SH) and domestic hot water (DHW) systems are controlled both centrally and locally. The central control is based on a heating curve, which describes the supply water temperature as a function of the outdoor temperature. Local adjustment is based on thermostatic radiator valves or room thermostats, keeping the indoor temperature set by adjusting water flow through radiators. Interacting with dynamic ventilation and transmission heat losses, these controls are responsible for the final indoor temperature and resulting space heating energy consumption, i.e. they can be targeted to reduce energy consumption. While both control levers have been shown to influence heat energy consumption (Cholewa et al., 2022; Werner-Juszczuk & Siuta-Olcha, 2024), the efficiency of thermostatic radiator valves and thermostats depends on the collaboration of occupants (Aragon et al., 2022; Bandurski et al., 2023; Bruce-Konuah et al., 2018; Lomas et al., 2018). The fundamental issue seems to be how these local controls are used by occupants and if they are suitable for their needs to control indoor thermal environment (Aragon et al., 2022; Bandurski et al., 2023).

Domestic hot water (DHW) usage is responsible for the energy used for heating up the required amount of water to the target temperature. This energy demand is primarily influenced by the temperature of supplied-to-building cold water, set temperature of DHW and water consumption. In central DHW systems, the circulation loop ensures immediate hot water availability, reducing water waste but significantly increasing heat consumption. Optimizing DHW systems therefore offers the potential to reduce energy consumption (by changing the temperature set-point) and water consumption (by changing the circulation loop flow) in buildings (Hofer et al., 2023; Klimczak et al., 2022; Van Thillo et al., 2022), while not impacting comfort (Huang et al., 2020). The influence of space heating (SH) and domestic hot water (DHW) control on energy consumption has been shown to effectively reduce energy consumption (Benakopoulos et al., 2021; Cholewa et al., 2019, 2022; Werner-Juszczuk & Siuta-Olcha, 2024). However, the scope of prior research is constrained to individual systems and their energy aspects, often based solely on simulation methods. Further, focusing on a single system neglects possible interaction between the use of SH, DHW and plug load systems. To overcome this shortcoming, it is necessary to analyze energy and water consumption on building level, not merely on system level.

Beyond this technical view, changing energy consumption in buildings cannot be tackled without taking into account possible effects on occupants and their behavior, which in the end shapes the final consumption (Van Den Brom et al., 2019). Technical interventions do not always lead to reduction in energy consumption; sometimes, gains in energy are offset by higher usage from occupants, termed a “rebound effect” (Sorrell et al., 2009, 2020). Rebound effects have long existed and have in the literature been shown for example after thermal retrofitting (Bardsley et al., 2019) or improvement of heating systems efficiency (Guerra Santin, 2013). Another source of subsequent behavior change following technical adjustments – such as changes to heating temperatures or hot water flows - might be discomfort, leading to an active override of the technical change. In one study, moderate adjustments to thermostat defaults in an office setting effectively reduced energy consumption, while drastic changes led the office workers to override the setpoints and consume more energy than before (Brown et al., 2013). It is therefore crucial to consider people’s experiences with technical interventions: here, research on interventions which target behavior change can inform and complement interventions which have a technical focus.

To summarize, research so far has either investigated technical building interventions and their direct outcome, or behavior change interventions and their impact on consumption. We aim to align these two perspectives to find out which kind of interventions are most promising to reduce energy consumption in space heating and domestic hot water, testing not only a psychologically informed intervention but also different technical interventions which consider rebound and spillover effects in different energy vectors. Additionally, we enrich our behavioral assessment with a survey, measuring the social impact of these interventions. By considering aspects such as acceptance of technical interventions, perceived social norms and collective efficacy beliefs as proxies of social identity processes, we can assess the underlying psychological changes which occur through being part of heating interventions as a form of collective action. Together with behavioral measures, we triangulate our different data sources to gain a fuller picture on the impact of heating and hot water interventions, demonstrating that strategies based solely on technical interventions in buildings risk underestimating the influence of human behaviors and social dynamics. Combining an assessment of direct impact on energy consumption, behavioral spillover and social impact measurement further contributes substantially to external validity (Hageman, 2008; Toresen Lokdam et al., 2021) and helps overcome limitations of single data sources (Sovacool et al., 2018), in order to create scalable strategies that effectively address both individual actions and systemic factors for energy savings.

### 2.4. Research questions

Overall, our study aims to identify and analyze the impacts of different interventions targeting space heating (SH) and domestic hot water (DHW) systems within a university dormitory setting. By employing a mixed methods approach we triangulate energy consumption and survey data, and aim to address several key research questions.

First, we investigate whether heating interventions, including both active behavioral requests and technical adjustments to space heating and domestic hot water systems, lead to a measurable reduction in energy consumption within the targeted domains. This aspect of the research focuses on understanding the direct impact of each intervention type on the consumption of space heating, electricity, and hot water.

RQ1 Do heating and hot water interventions reduce energy use in their target domains, without creating rebound effects?

- RQ1a: Do heating interventions reduce space heating consumption?
- RQ1b: Do hot water interventions reduce hot water consumption?

- RQ1c: Do any interventions increase energy use in their target domain (rebound effect)?

Second, the study explores the potential for behavioral spillover effects. Specifically, we examine whether interventions lead to changes in energy usage in other areas, such as electricity consumption. This inquiry is essential for understanding whether engaging with interventions in one domain may inadvertently influence behavior in another, either positively or negatively.

RQ2 Do heating and hot water interventions affect energy use in untargeted domains, e.g. electricity (spillover effect)?

Finally, we explore the social implications of participating in these sustainable energy interventions. The study examines whether involvement in the interventions fosters positive or negative social outcomes, such as a positive or negative attitude towards the interventions, increased or decreased support for pro-environmental policies, enhanced collective efficacy beliefs among students, or a shift in perceived social norms.

RQ3: Does participation in the interventions lead to self-reported positive social outcomes?

# 3. Method

## 3.1. Situational context

The study was conducted in collaboration with the student dormitories of a Polish university. For our field trial, we picked three low-rise dormitory buildings (A, B and C, as shown in Figure 1), which are all situated within the same area of the campus and provide shared living spaces for both national and international students. The trial buildings were selected for their similar shape and characteristics, especially for having a nearly identical glazed-to-wall ratio, ensuring comparable thermal performance. Each building is equipped with radiator heating, supplied by the district heating network via a substation. This heating substation is fully automated and integrated with the university's Building Management System, also supplying domestic hot water and facilitating a hot water circuit. The system is slightly different in Dormitory B, which underwent modernization a few years ago. This upgrade included the installation of room temperature sensors and electrothermal radiator valves, increasing the automation of the facility. Additionally, the ventilation system was upgraded to a mechanical supply and exhaust system with heat recovery and fully integrated into the building automation system. For more technical information on the dormitories, see Supplementary Materials (SM1).

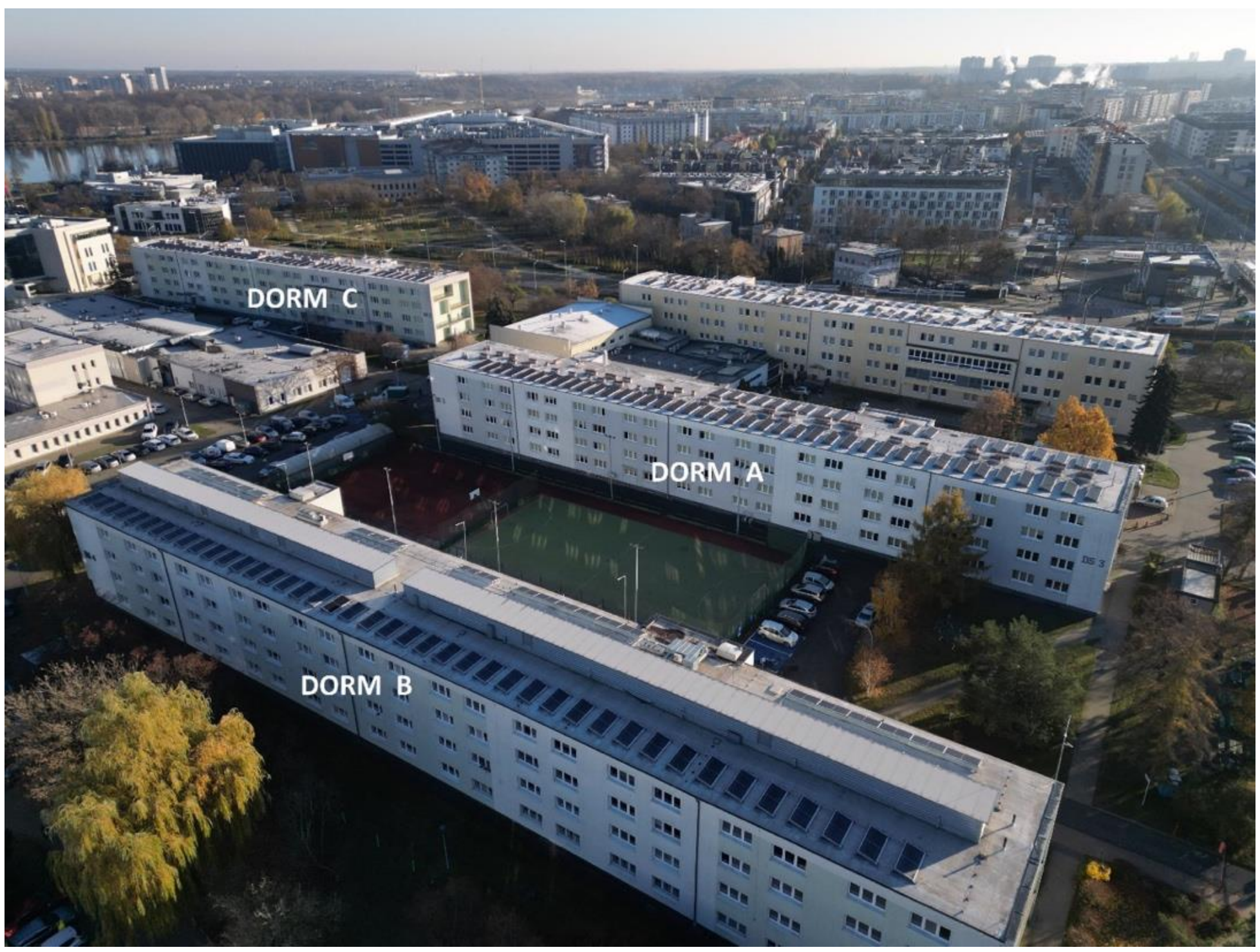

*Figure 1: This figure shows a view of the dormitories included in the trial (A, B and C).*

To assess the impact of our interventions on energy consumption and behavior, we employed a quasi-experimental design. Two dormitories (A, B) were selected to receive targeted interventions aimed at influencing space heating (SH) and domestic hot water usage (DHW), while a third dormitory (C) was designated as a control group. This design allows for a more rigorous examination of the expected causal effects of the interventions by comparing the change in outcome variables between the intervention groups and the control group.

**3.2. Research design and participants**

The study was preregistered prior to data collection and ethical approval was obtained from the Institutional Ethical Review Board prior to the commencement of the study. In a pre-post and control design, we compared energy consumption measurements before and during the interventions, in the control (C) and the two intervention dormitories (A, B). Interventions (described in detail in Table 1) in both dormitories were implemented subsequentially: First, an active request to students living in dormitories A & B was sent out, asking for a reduction in space heating ("Active SH intervention"). After that, a technical space heating intervention through the system ("Passive SH intervention") was implemented in both dormitories, followed by two different interventions targeting first only the circulation pump of domestic hot water ("DHW pump intervention"); and later additionally the domestic hot water temperature ("DHW temperature & pump intervention") in dormitories A & B. Energy consumption was measured on building level daily for all energy vectors (hot water, space heating and electricity) in all dormitories (A, B & C). Additionally, we complemented the energy consumption data with a within-between subject survey, assessing social impact measures through a questionnaire sent to all students living in dormitories A, B and C before interventions started (T1), and after the space heating interventions (T2). Considering potential technical and social differences in the dormitories at baseline, all analyses focus on the difference-in-difference, i.e. in how far the interventions influenced the priorly measured baseline patterns within dormitories.

Each dormitory houses approximately N = 250 occupants. For the within-between subject design in the survey, we gathered N = 74 responses from participants (students from dorm A, B and C) at both time points. Survey participants were recruited via flyer and e-mail, and were compensated for their participation with a local shop voucher. All interventions were implemented per dormitory for A and B, while no interventions were implemented in dormitory C. Occupancy in the dormitories differed only marginally throughout the trial period and was considered as a control variable in all analyses. Figure 2 illustrates the procedure of the trial.

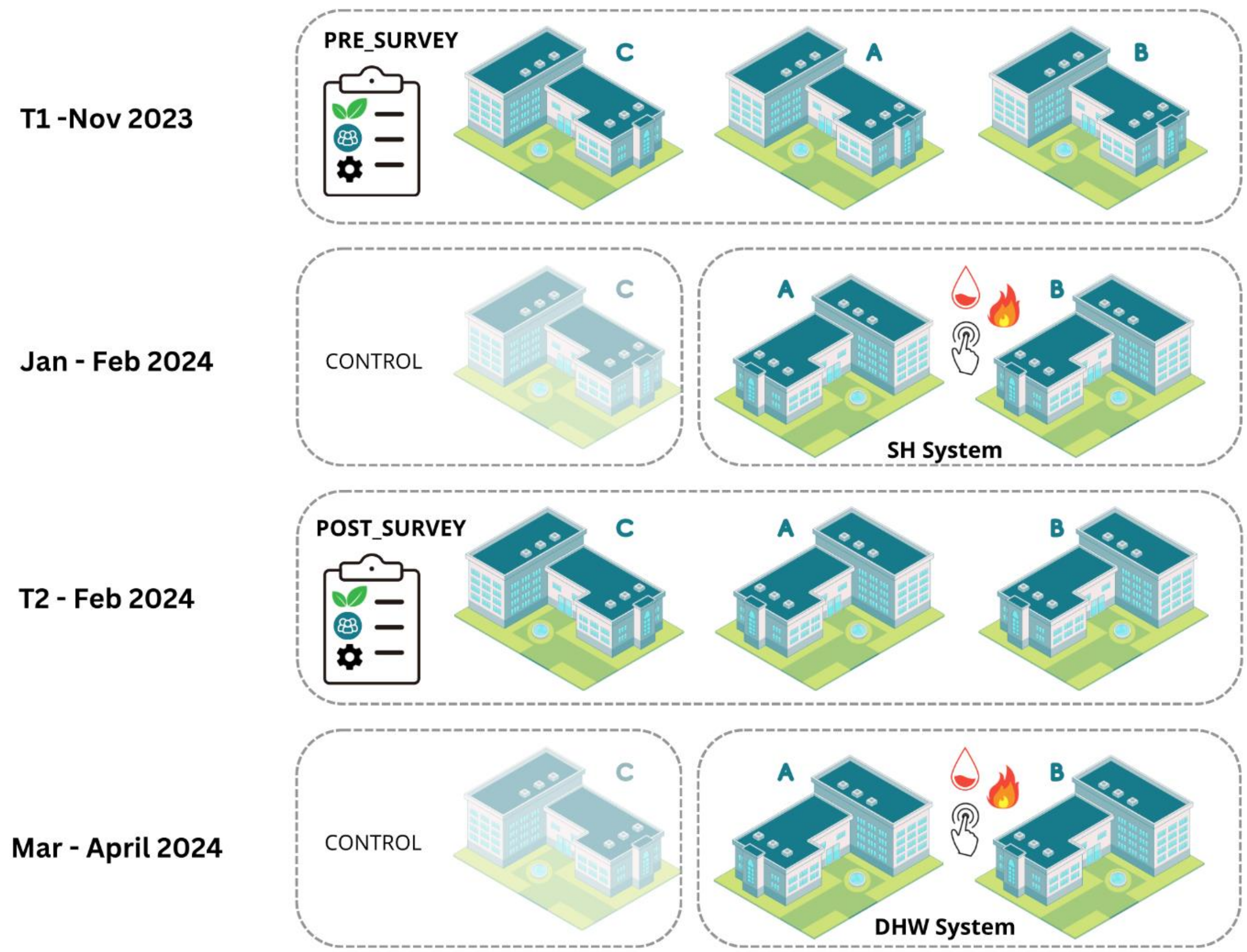

*Figure 2: Procedure of the trial, including survey times and indicating control vs. intervention group.*

### 3.3. Materials

**3.3.1 Interventions.** Table 1 gives an overview over the interventions implemented in intervention dormitories A and B. Overall, we had four key intervention periods: One with an active request to reduce space heating (SH) using a psychologically informed intervention (1), one intervention targeting space heating (SH) through lowering the heating curve (2), a third intervention targeting the circulation pump of domestic hot water (Circulation$^{DWH}$) (3) and the last one, which targeted both circulation pump of domestic hot water (Circulation$^{DHW}$) and domestic hot water temperature (DHW) (4). The table displays the number of interventions, the dates in which the intervention was implemented, i.e. the intervention period, the intervention name and how it was implemented in the respective intervention dormitory. For each of these interventions, there are two energy consumption related research hypotheses: Based on our RQ1 ("Do heating and hot water interventions reduce energy use in their target domains, without creating rebound effects?"), we focus on the targeted system, i.e. if interventions 1 and 2 decrease SH consumption, and interventions 3 and 4 decrease DHW and Circulation$^{DHW}$ consumption. Further, in line with our RQ2 ("Do heating and hot water interventions affect energy use in untargeted domains, e.g. electricity (spillover effect)?"), we control for any increase in energy consumption in other domains. Here, we assume that interventions 1 and 2 do not create negative spillover effects in electricity energy, domestic hot water energy and volume consumption, while interventions 3 and 4 should not lead to negative spillover effects in energy consumption of SH or electricity.

*Table 1: Interventions overview.*

| No. | Date | Name | Dorm A | Dorm B |
|---|---|---|---|---|
| 0 | 04.12.23-15.01.24 | // | *Reference Period* | |
| 1 | 16.01.24-22.01.24 | **Active SH intervention** | *Request to residents:* lower the setting by the end of the week | *Request to residents:* lower the setting by the end of the week |
| 2 | 22.01.24-13.02.24 | **Passive SH intervention** | *Technical intervention*:<br>- Reduce SH heating curve by 3°C | *Technical intervention*:<br>- Reduce SH heating curve by 3°C<br>- Ventilation temperature at 19°C<br>- Heat recovery in ventilation unlimited<br>- Reduction of internal temperature set point by 1ºC |
| 3 | 06.03.24-11.03.24 | **DHW pump intervention** | *Technical intervention*:<br>- Circulation pump OFF if VDHW>0.6 m3/h and ON if VDHW<0.5 m3/h | *Technical intervention*:<br>- Circulation pump OFF if VDHW>0.7 m3/h and ON if VDHW<0.6 m3/h |
| 4 | 11.03.24-20.03.24[3] | **DHW temp & pump intervention** | *Technical intervention*:<br>- Circulation pump OFF if VDHW>0.6 m3/h and ON if VDHW<0.5 m3/h<br>- DHW set temperature 45ºC | *Technical intervention*:<br>- Circulation pump OFF if VDHW>0.7 m3/h and ON if VDHW<0.6 m3/h<br>DHW set temperature 45ºC |

**3.3.2 Framing participation as collective action:** Students in the intervention dormitories were informed about the technical interventions within their student housing. All communication (active request, information) framed the interventions and effort as a collective action to reduce carbon emissions. As part of the first space heating intervention, students (intervention group) were requested to actively contribute to the heating reduction by turning down their thermostats

[3] Please note that due to a complaint, the DHW temperature & pump intervention was stopped in Dorm A after three days. Therefore, some data in the final analysis for Dorm A will not be available, labeled ‘NA’. To account for this, we also split the fourth intervention into two periods, 4a for both dormitories, and 4b for Dormitory B only.

(intervention 1, dormitory A and B). Additionally, in the weeks after, students in the intervention dormitories got multiple e-mails informing them about their participation and the collective action to reduce $CO_2$ emissions.[4]

**3.3.3 Dependent variables.** The key dependent variables to assess energy consumption behavior are the amount of heat consumption (space heating, in kWh), energy consumed for hot water and hot water circulation (hot water, in GJ), water volume used (hot water, in $m^3$) and electricity energy consumption (in kWh), all measured on building level. These were extracted from the available system data from the local Building Management System. The key dependent variables for social impact were the attitude towards the energy saving system implemented in dormitories, perceived social norms and pro-environmental policy support as a measure of spillover, estimated with the survey (see 3.3.4).

**3.3.4 Survey.** For the survey, we collected all data using SosciSurvey – the questionnaire is available in the open access repository and in the additional supplementary material. Unless otherwise specified, change scores were calculated as the absolute arithmetic difference between second-survey scores and first-survey scores.

***Attitude.*** In the survey, students were first introduced to the concept of *The Energy Saving Control Strategy,* aiming at lowering energy consumption and $CO_2$ emissions through intervening with the heating and hot water system. Afterwards, we assessed their attitude towards the strategy through their agreement to statements on a 7-Point Likert scale, for example including "*I think the energy saving control strategy in my dormitory is a good idea*.". Further, they were asked to indicate their feelings, ranging from very negative to very positive towards implementation of the strategy and people who support it. Cronbach's alpha for the four attitude items was **α** = .79.

[4] An example of such an email can be found in the Supplementary Materials (SM2)

***Student identification and collective efficacy***. For student identity, we asked for belongingness and pride of students to be part of the student community with two items, resulting in a Cronbach's alpha of **α** = .83. For collective efficacy beliefs, we assessed both the belief to be able to make a difference, e.g. "*I think that we, as student community of the [Polish university], can make a difference by heating more sustainably in the long run*" and the belief to be able to manage more sustainable heating "*I think that we, as student community of the [Polish university], can manage to heat in a more sustainable way.*", based on (Jugert et al., 2016). The four-item scale had a Cronbach's alpha of **α** = .90.

***Social norms.*** We aimed to assess both descriptive and injunctive student (i.e. peer, in-group) norms and additionally injunctive authority norms. We therefore asked students whether they believe that most other students will or will not accept the strategy (descriptive norms, two items, 7-point scale), and whether other students (and the university authorities) expect them to accept it (injunctive norms, two items each group, 7-point-scale). Descriptive norms as well as injunctive norms for both students and authority were all assessed with each two items, and Cronbach's alpha were **α** = .83, **α** = .72 and **α** = .80 respectively.

For all constructs, we assessed to what extent they agreed with listed statements on a 7-point Likert scale between strongly disagree (1) and strongly agree (7).

Further, all participants indicated information on their demographic data, their political orientation, their time lived in the dormitory and their plans to stay in Poland.

### 3.4. Analysis

First, the overall heat energy demand data were cleaned from the influence of outdoor temperature. For this, to facilitate the evaluation of interventions' impacts on the facility's energy intensity, the relationship between heat demand and outdoor temperature was analyzed. Historical data from the period between December 4, 2023, and January 15, 2024, were utilized to establish this relationship using a linear regression. Before the analysis, several assumptions were made: only periods where solar radiation was below 100 W/m² were considered, to eliminate the influence of solar energy, and only data points with an outdoor temperature below 15 °C were included. The average heat demand for each outdoor temperature was then calculated to the nearest degree Celsius. Based on these averages, the following regression equations were derived for each dormitory:

$$f_A(t_{out}) = -3.10 \cdot t_{out} + 81.3 \text{ [kW]} \qquad (1)$$

$$f_B(t_{out}) = -4.36 \cdot t_{out} + 78.0 \text{ [kW]} \qquad (2)$$

$$f_C(t_{out}) = -4.01 \cdot t_{out} + 84.7 \text{ [kW]} \qquad (3)$$

where: $t_{out}$ – outdoor temperature [°C]

The Figure 3 below illustrates the graphical representation of equations 1, 2 and 3, with the use of dashed lines.

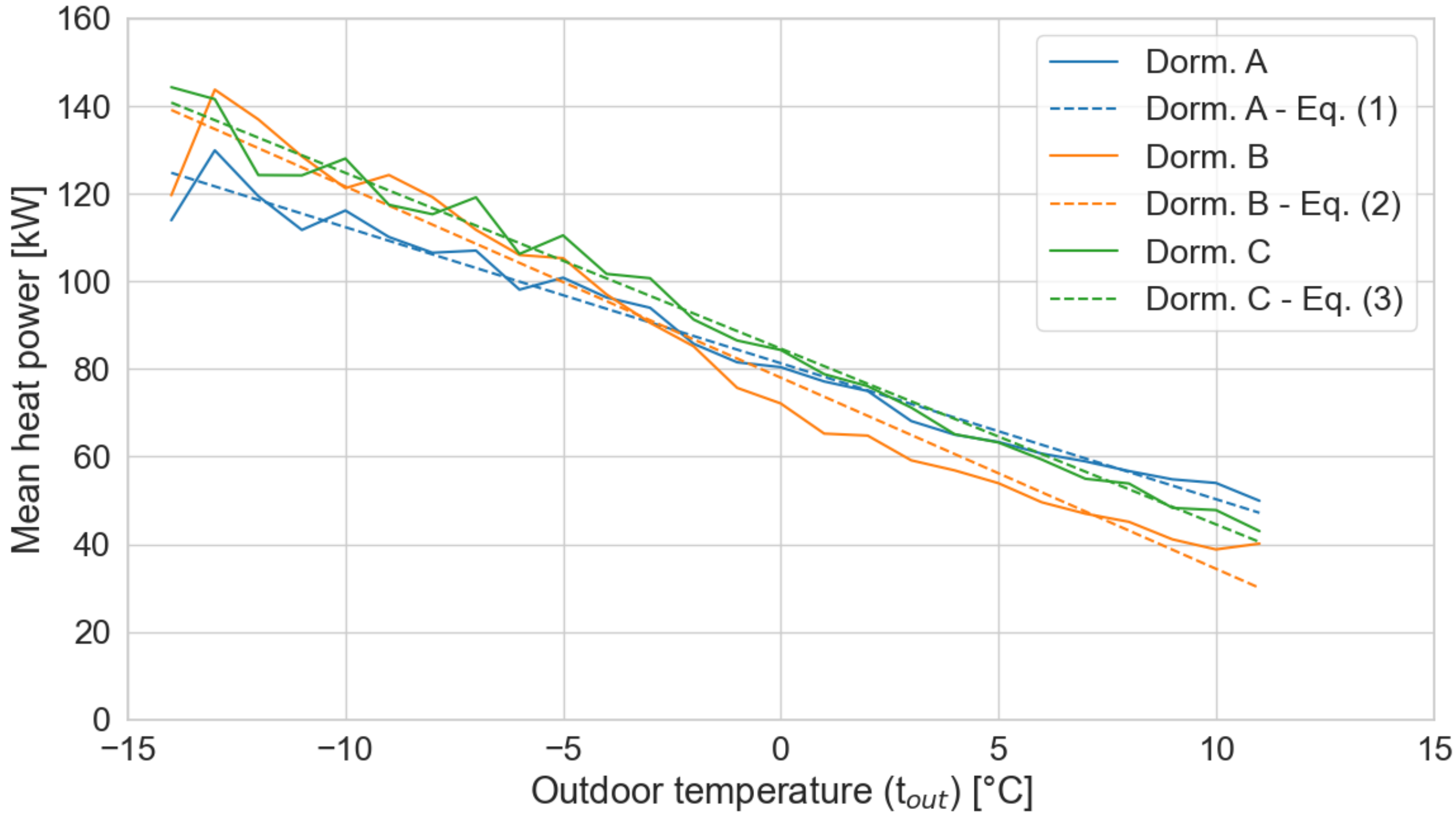

*Figure 3: Graphical representation of relation between heat demand and outdoor temperature.*

Further, two metrics are used to analyze intervention results for energy consumption. The main metrics capture the intervention impact (4 and 5), which describe how much the total energy consumption changed after an intervention between the tested dormitory and the reference dormitory in comparison to the pre-intervention period (eq (4), for Circulation$^{DHW}$, DHW, electricity).

$$II = \frac{cons_{int.j,sys,u,D}}{cons_{int.j,sys,u,Dref}} - \frac{cons_{int.0,sys,u,D}}{cons_{int.0,sys,u,Dref}} \qquad (4)$$

where:

II - intervention impact on energy consumption
j - intervention number (1-4)
sys - considered system (SH, $C^{DHW}$, DHW, EE)
u - consumed utilities (heat, electricity or water)
D - dormitory name (DSA, DSB)
$D_{ref}$ - reference dormitory name (DSC)
cons - energy consumption

int.j - period of jth intervention, e.g. 1 is Jan 16 – Jan 22
int.0 –period no intervention (reference): Dec 4 –Jan 16

or how much the consumption changed in comparison to the simulated energy consumption for the intervention period (eq. (5), for SH):

$$II_{j,sys,u,D} = \frac{cons_{int.j,sys,u,D}}{cons_{int.j,sys,u,Dref}} - \frac{modeled\ cons_{int.j,sys,u,D}}{modeled\ cons_{int.j,sys,u,Dref}} \quad (5)$$

where:

modeled cons - energy consumption for SH calculated based on a linear model calibrated based on reference period (see Table 1)

Simulated energy consumption of buildings for SH is based on a model calibrated on data from the pre-intervention period, with outdoor temperature as model input. Occupancy numbers could influence consumption in a direct way, e.g. more occupants imply a higher DHW and electricity consumption. However, for SH, the situation is more complex: on one hand, the presence of occupants can cause an operation mode for SH or the occupants could open the windows (increasing SH demand), but on the other hand, more occupants can induce internal heat gains, which decreases SH demand. Therefore, the occupants' number is not included in the main metric but is controlled independently to show the descriptive results in a more explicit way. Occupant changes metric is analogous to *ii*:

$$OCCCH_{j,D} = \frac{occ_{int.j,D}}{occ_{int.j,Dref}} - \frac{occ_{int.0,D}}{occ_{int.0,Dref}} \quad (2)$$

where:
OCCCH - change in occupancy
occ - occupant number

To test the statistical effect of the interventions on energy consumption both on the target variables of the respective intervention (i.e. intervention 1 and 2 on space heating, and 3 and 4 on energy consumption by hot water), as well as on potential spillover variables (e.g. amount of hot water consumed, electricity), we employ a linear regression model for each dependent variable separately. By including intervention time (i.e. the intervention vs. the pre-intervention period) and dormitory (i.e. dormitory A and B = intervention vs. dormitory C = Control) into this regression model, we control for individual differences between dormitories (between-subject factors) and assess changes over time (within-subject factors). The main focus is the interaction of interventions with dormitories, demonstrating if an intervention had a significant effect on energy consumption patterns across different dormitories. Additionally, we control for other covariates (occupancy numbers, weekday, outside temperature, hour of the day). An example of one regression model we ran in R is provided below for space heating:

$$kWh_it = \beta_0 + \beta_1 \cdot intervention_{\mathrm{i}} + \beta_2 \cdot dorm_{\mathrm{i}} + \beta_3 \cdot (intervention_{\mathrm{i}} \times dorm_{\mathrm{i}}) + \beta_4 \cdot occupancy_it + \beta_5 \cdot weekday_t + \beta_6 \cdot hour_t + \beta_7 \cdot outsideTemp_t + \varepsilon_it$$

## 4. Results

All statistical analyses were conducted using R (version 4.4.3; R Core Team, 2024) and Microsoft Excel.[5] To understand the overall impact of the interventions on energy consumption, we first tested the effects of all interventions on all dependent variables, both for the targeted energy domains of this intervention (e.g. Active and Passive SH interventions on space heating kWh consumption), and for potential spillover effects (e.g. Active and Passive SH interventions on hot water GJ usage). Table 2 shows both the results of the descriptive analysis, describing the change in ratios for the different energy domains, and the statistical outcomes of the regression results for interactions between dormitories and interventions.

[5] Data files and analysis code are openly available under the following link: https://osf.io/94dxr/?view_only=d79dea2e6d1d44d2a9ac8da5bcf17080

*Table 2: Main results, structured along intervention and outcome variable.*

| Outcome | Method | Active SH intervention | Passive SH intervention | DHW pump intervention | DHW temperature & pump intervention |
|---|---|---|---|---|---|
| **Space heating (kWh)** | descriptive | $II_{1,SH,h,DA}$ = -1%<br>$II_{1,SH,h,DB}$ = -7% | $II_{2,SH,h,DA}$ = -13%<br>$II_{2,SH,h,DB}$ = -21% | $II_{3,SH,h,DA}$ = +2%<br>$II_{3,SH,h,DB}$ = 0% | $II_{4,SH,h,DA}$ = NA<br>$II_{4,SH,h,DB}$ = -4% |
| | inference statistics | Significant negative interaction of intervention 1 with dormitory A (ß = -4.38, $p$ <.001) and dormitory B (ß = -5.23, $p$ <.001) | Significant negative interaction of intervention 2 with dormitory A (ß = -4.53, $p$ <.001) and dormitory B (ß = -10.05, $p$ <.001) | Significant positive interaction with dormitory A (ß = 3.77, $p$ = .016), no significant interaction with dormitory B ($p$ = .634) | Significant positive interaction for dormitory A for intervention 4a (ß = 7.67, $p$ = <.001), no interactions for dormitory B for either 4a ($p$ = .827) or 4b ($p$ = .826) |
| **DHW total (energy in GJ;** includes heating DHW amount in $m^3$ to target temperature in °C) | descriptive | $II_{1,DHWtot,h,DA}$ = -3%<br>$II_{1,DHWtot,h,DB}$ = +3% | $II_{2,DHWtot,h,DA}$ = -3%<br>$II_{2,DHWtot,h,DB}$ = +7% | $II_{3,DHWtot,h,DA}$ = -1%<br>$II_{3,DHWtot,h,DB}$ = +9% | $II_{4,DHWtot,h,DA}$ = NA<br>$II_{4,DHWtot,h,DB}$ = +5% |
| | inference statistics | Small significant negative interaction effect of intervention 1 with dormitory A (ß = -0.11, $p$ = .027), no significant effect with B ($p$ = .061) | Small significant negative interaction effect of intervention 2 with dormitory A (ß = -0.13, $p$ < .001), positive interaction effect of intervention 2 with dormitory B (ß = 0.15, $p$ < .001) | No interaction with dormitory A ($p$ = .131), positive interaction effect of intervention 3 with dormitory B (ß = 0.239, $p$ < .001) | No significant interaction with both dormitory A for 4a ($p$ = .327) and dormitory B for 4a ($p$ = .903) or 4b ($p$ = .180). |
| **DHW amount (water $m^3$)** | descriptive | $II_{1,DHW,w,DA}$ = -3%<br>$II_{1,DHW,w,DB}$ = +2% | $II_{2,DHW,w,DA}$ = -7%<br>$II_{2,DHW,w,DB}$ = +4% | $II_{3,DHW,w,DA}$ = +12%<br>$II_{3,DHW,w,DB}$ = +17% | $II_{4,DHW,w,DA}$ = NA<br>$II_{4,DHW,w,DB}$ = +20% |
| | inference statistics | Small significant negative interaction effect of intervention 1 with dormitory A (ß = - | Significant negative interaction effect of intervention 2 with dormitory A (ß = -1.09, $p$ | Significant positive interaction of intervention 3 with dormitory A (ß = 1.41, $p$ = .002), significant | Significant positive interaction of intervention 4a with dormitory A (ß = 0.96, $p$ = .044), significant positive interaction of intervention 4a with |
| | | 1.09, $p < .001$), no effect with dormitory B ($p = .118$) | $< .001$), positive interaction effect with dormitory B (ß = 0.793, $p < .001$) | positive interaction with dormitory B (ß = 2.46, $p < .001$) | dormitory B (ß = 2.26, $p = <.001$) and 4b with dormitory B (ß = 3.6, $p < .001$) |
| **Electricity (kWh)** | descriptive | $II_{1,EE,e,DA} =$ -3%<br>$II_{1,EE,e,DB} =$ +3% | $II_{2,EE,e,DA} =$ -3%<br>$II_{2,EE,e,DB} =$ +6% | $II_{3,EE,e,DA} =$ -5%<br>$II_{3,EE,e,DB} =$ +3% | $II_{4,EE,e,DA} =$ NA<br>$II_{4,EE,e,DB} =$ +7% |
| | inference statistics | No significant interaction with both dormitory A ($p = .721$) and dormitory B ($p = .213$) | Significant positive interaction with dormitory B (ß = 1.34, $p < .001$), no interaction with dormitory A ($p = .186$) | Significant negative interaction with dormitory A (ß = -1.66, $p = .024$), no interaction with dormitory B ($p = .394$) | Significant positive interaction of intervention 4a with dormitory B (ß = 2.01, $p = .015$) and 4b with dormitory B (ß = 1.60, $p = .008$), no interaction of 4a with dormitory A ($p = .692$) |
| Occupants (number) | descriptive | $OCCCH_{1,DA} =$ -1%<br>$OCCCH_{1,DB} =$ -2% | $OCCCH_{2,DA} =$ -3%<br>$OCCCH_{2,DB} =$ -1% | $OCCCH_{3,DA} =$ -5%<br>$OCCCH_{3,DB} =$ +3% | $OCCCH_{4,DA} =$ -6%<br>$OCCCH_{4,DB} =$ +4% |

### 4.1. RQ 1: Targeted energy impact

Referring to our first research question, i.e. whether the interventions led to an actual reduction in energy consumption in the targeted energy domain, we focus on space heating (in kWh) for the Active and Passive SH interventions and on total energy demand for domestic hot water (in GJ) for both of the DHW interventions. Overall, we see a reduction of energy consumption in space heating for both the active request and the technical space heating interventions. During both intervention periods, we find a significant negative interaction effect, i.e. a decrease in space heating kwh when compared to the control dormitory. For the DHW interventions, we find a different pattern. Here, we see no significant effects, i.e. the DHW pump intervention and the HW pump & temperature intervention did not significantly interact with hot water GJ consumption in dormitories. Notably, for the DHW pump intervention, we even see a positive effect of the interaction, i.e. an increase in hot water GJ consumption for dormitory B, indicating a rebound effect.

### 4.2. RQ 2: Spillover effects

For our second research question, we investigated behavioral spillover effects. For the Active SH intervention, we see no significant spillover effects in our data. For the Passive SH intervention, we find a small significant increase in electricity kWh use and DWH use for dormitory B only. However, overall, both of the space heating interventions were very effective in reducing total energy consumption significantly. Again, we see different patterns for the two hot water interventions: the DHW pump intervention led to significant negative spillover effects, i.e. an increase in both energy demand for space heating in kWh and in water volume used in $m^3$. The DHW temperature & pump intervention, next to already leading to a complaint and therefore drop out of Dormitory A, again significantly increased water volume $m^3$ used in both intervention dormitories, demonstrating a negative spillover effect.

### 4.3. RQ 3: Social impact

For our third research question, we assessed the social impact created by the interventions. We collected a final sample of N = 74 students who had participated in both surveys (pre- and post SH interventions). Of these, N = 19 students were living in control dormitory C, N = 24 in dormitory A and N = 31 in dormitory B. The participants' ages ranged from 17 to 23, with a mean of age of 21.16 (SD =2.67). Most participants were male (69%) compared to female (24%) or other (6%). There were no significant differences between dorms in age ($p$ =.275) or gender ($p$ = .072).

We assessed the baseline for the main dependent variables (attitude, collective efficacy beliefs, social norms injunctive and descriptive) at T1 to check for baseline differences between dorms. Overall, there were no significant differences for variables at baseline T1 between dorms. To account for the baseline in each dorm at T1, we will interpret the difference-in-difference changes between T1 and T2 in dorms, using interactions between dorms and point in time as an indicator for an effect of participating in the interventions. The following Table 3 demonstrated this difference-in-difference for intervention vs. control. The values displayed show the change scores as the arithmetic difference between T2 scores and T1 scores.

*Table 3: Difference scores for social impact assessment.*

| **Variable** | **Control** | **Intervention** | ***p-value*** |
|---|---|---|---|
| Attitude difference | -0.09 | 0.02 | .694 |
| Descriptive norms difference | 0.35 | 0.40 | .894 |
| Injunctive norms difference | -0.75 | 0.14 | .021* |
| Authority norms difference | -0.68 | 0.08 | .062 |
| Collective efficacy difference | -0.19 | -0.34 | .517 |
| Social identification difference | -0.38 | -0.23 | .642 |
| Policy support difference | 0.03 | -0.11 | .502 |
| CGB difference | -0.13 | -0.12 | .957 |

* = p <.05, ** = p <.001

Descriptively, we see only very small difference in attitude in the dormitories, and the biggest difference through a decrease in injunctive norm perception (both related to other students and authorities) for Dorm C while having an increase on these variables in the intervention dormitories. Descriptive norms seem to increase slightly for both intervention and control group at T2. We see a small decrease in pro-environmental policy support in the intervention dorms. For social identification, CGB in all dormitories and collective efficacy, we see similar small decreases at T2 for both intervention and control. Looking further into this interaction between *intervention group* and *time* in an ANOVA for each of the dependent variables, we see no significant interactions between time and intervention, i.e. no difference in difference for people experiencing the intervention. However, this might also be because of our small sample size. Using an a priori analysis with G-power, we determined that for an ANOVA with repeated measures, within-between interaction with two groups (intervention vs. none) and two measurement time points (pre- and post), with a non-sphericity correction of m / (m-1) = 1, and an assumption of a small effect (.15), a sample size of 196 participants would have been needed to find an effect (alpha = .05, power = .95). As we did not manage to meet this number of participants, we exploratorily used a linear regression model to predict the difference score (calculated as the arithmetic difference between second-survey scores and first-survey scores) by intervention group, as a post-hoc power analysis calculated with G-power indicated that with our sample of N = 74, a linear regression with one predictor (alpha = .05) could reach a power of $>.95$. We do not find significant effects for pro-environmental policy support ($p = .520$), attitude ($p = .649$, descriptive social norms ($p = .894$) and injunctive authority norms ($p = .062$), but there is a significant effect of the intervention on injunctive social norms (ß = 0.89, $p = .021$). All test results of difference scores can be found as well in Table 3.

Figure 4 visualizes the interaction between intervention group (control vs. intervention dorms) and time (T1 = pre-intervention; T2 = post-intervention) for injunctive student norms, demonstrating the small significant effect: while the injunctive norm perception in the intervention group slightly rises, it decreases in the control group. This may be due to students in the control group recognizing they were not included in the intervention, which could have lowered their perception of social norms, specifically, the sense that others expect them to contribute. In contrast, norm perceptions appear to increase when individuals are actively involved in, and reminded of, the collective energy action.

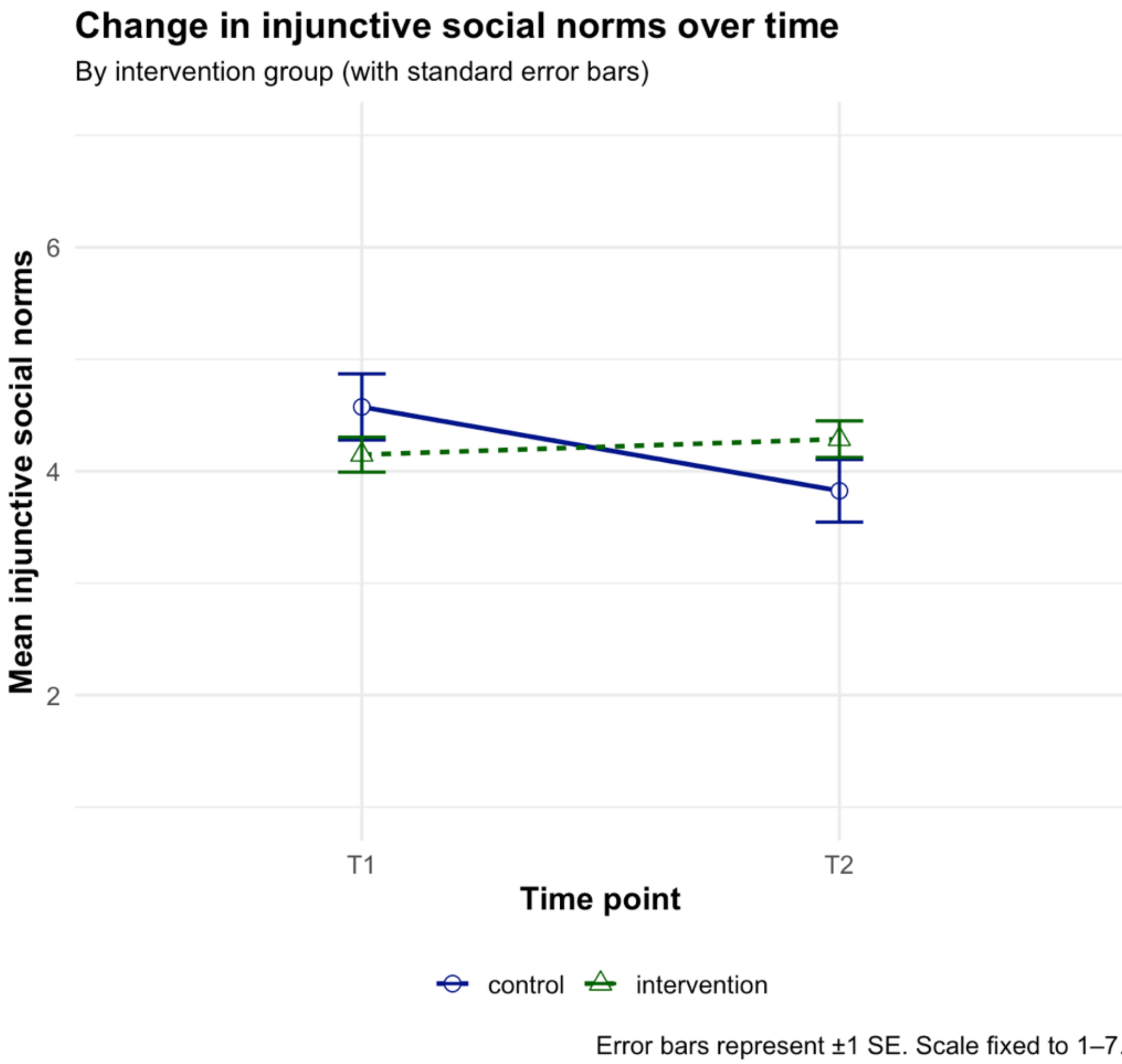

*Figure 4: Effect of intervention vs control over time for injunctive student norms.*

## 5. Discussion

Overall, our study shows that while various interventions in heating and hot water systems have the potential to reduce overall energy consumption, their effectiveness ultimately depends on people's behavior. Our first research questions focused on the directly targeted energy impact, asking if the intervention was successful or created a rebound effect. As a result, we found that the most promising interventions were an active request to lower the thermostat point and an automatic intervention targeting the set temperature of the heating curve. Both of these led to a measurable reduction in space heating consumption. In contrast, interventions targeting the circulation of domestic hot water and the domestic hot water temperature setpoint did not result in reduced energy use. Notably, the DHW pump intervention even demonstrated a small rebound effect, increasing hot water energy consumption.

For our second research question, we examined spillover behaviors into other energy consumption domains. We observed that the DHW interventions led to negative spillover effects — specifically, an increase in water volume used (and therefore energy used for hot water) and an increase in space heating consumption. In contrast, no spillover effects were found for either of the space heating interventions.

Our third research question asked for the social impact of framing the interventions collectively. Our findings suggest that norm perceptions changed based on whether participants were actively part of the collective energy action and reminded of it. Other than that, we did not find any significant effects, including no negative effects on attitude towards the system.

Overall, one core finding of our research is that technical interventions must be considered carefully in terms of unintended consequences when accounting for behavior: reduction in temperature setpoints (e.g. in hot water) might not always lead to actual reduction of energy consumption if it triggers negative spillover effects and resistance. A possible explanation why the hot water interventions produced such negative effects is that these kinds of interventions are perceived as more drastic, as they may have a more noticeable impact on comfort. Unlike space heating, which can be adjusted more subtly, interventions affecting domestic hot water are often harder for occupants to override or adapt to. Furthermore, it's plausible that after a period of ongoing interventions, students grew frustrated, contributing to the negative reception of these changes.

Both of our space heating interventions showed promising effectiveness. This is particularly relevant considering that there was no financial incentive involved in the first intervention, but just a request to reduce temperature for a specific time frame motivated by a collective, pro-environmental effort. Also, even when reducing temperature automatically and informing students about interventions in their heating systems, students' attitudes towards these intervention strategies did not decrease and we found no rebound effects or resistance. Although the social impact created through intervention participation was rather low, there are some trends that show the relevance of these actions for perceived social norms.

**Limitations and strengths**

Our study is not without limitations. The data available for analysis were aggregated at the dormitory level rather than the individual room level. This aggregation means we were unable to examine potential gender differences in comfort levels or assess whether specific rebound effects were more prominent among certain subgroups of students. However, due to privacy concerns and limited data availability, only aggregated data were available. The survey dropout rate and the relatively small sample size of respondents in our social impact survey present another limitation. With fewer responses, it becomes challenging to draw strong conclusions about the effects that the interventions had on our social impact outcome variables, especially considering the sample may not be fully representative of the full student population.

Despite these limitations our study also has notable strengths. Conducted as a field study, it reflects real-world conditions and actual energy behavior, providing robust evidence on time series consumption patterns rather than relying on simulations or hypothetical models. Testing interventions that combine behavioral and technical approaches is crucial to develop scalable and context-sensitive strategies to reduce heating and hot water energy consumption in households. These insights are relevant for building managers, policymakers and private sector actors as they can help to bridge the gap between technological solutions and human behaviors that drive their effectiveness.

## 6. Conclusion

Behavioral interventions can complement technical measures by addressing areas where interventions like price signals or efficiency upgrades alone may not achieve the desired results. They also help manage potential rebound effects, ensuring that any energy savings achieved are not offset by increased consumption in other domains.

Trialing these interventions in controlled, real-world settings is particularly important because it allows us to understand how technical systems, individual behavior, and social dynamics interact before scaling up (Berger et al., 2023). Additionally, the results contribute to the broader debate within the i-frame (individual-level) versus s-frame (system-level) interventions (Chater & Loewenstein, 2022). If interventions place too much emphasis on individual behavior without sufficient systemic support, they may not achieve meaningful or sustainable changes in the overall energy transition. The challenge lies in balancing these approaches to ensure interventions are both effective and acceptable to occupants over time.

## Ethical considerations

Ethical approval was obtained from the Institutional Ethical Review Board prior to the commencement of the study, with final approval on 05.12.2023.

## Informed consent

Informed consent was assessed in the study, i.e. participation in the pre/post evaluation survey is ensured through active recruitment and informed consent, as well as information about the voluntary nature of the survey. The technical innovation for the energy saving measure, about which the students are informed, is part of the technical infrastructure implemented and approved by the university. Informed consent for participating was waived by the review board, as an opt-out or opt-in design was technically not feasible. Instead, the accompanying research aimed to evaluate the extent to which the measure makes sense (both from a social and an energy consumption perspective).

## Supplementary Material

**SM1:** A detailed description of the student dormitories under study.

**Student dormitory C (control)**

- Number of student rooms: 118 rooms (96 3-bed rooms, 3 2-bed rooms).
- Hotel accommodation: 33 beds, including:
  - 5 compact single rooms (with access to shared sanitary facilities and kitchen),
  - 3 double rooms (with kitchenette, bathroom and TV),
  - 6 double rooms (with kitchenettes, bathrooms),
  - 2 double rooms (with kitchenettes, TV and access to a shared bathroom for guests),
  - 3 flats of 60 m² each (with kitchen, bathroom, hallway and TV).
- Additional facilities: Medical clinic and fitness club in the building.
- Full floor basement.

**Student dormitory A**

- Number of student rooms: 344 places (8 2-bed rooms, 103 3-bed rooms).
- Guest rooms:
  - 6 double rooms (with own kitchenette, bathroom and TV),
  - 7 single rooms (with access to kitchen and sanitary facilities).
- Full floor basement.

**Student dormitory B**

- Number of student rooms: 78 rooms (12 single rooms, 66 double rooms).

- Facilities: Rooms adapted to the needs of people with disabilities, including wheelchair users and the visually impaired.
- Additional information: General building renovation carried out in 2020 (installation of mechanical ventilation with heat recovery and implementation of Building Management System control); disabled accessible gym.
- Basement on a part of the floor (only space for district heating substation and distribution of heating).

The following table presents the characteristic parameters of the analysed dormitories (green shading highlights values that exhibit similarity).

| | | **Dorm C** | **Dorm A** | **Dorm B** |
|---|---|---|---|---|
| | Construction year | 1960 | 1964 | 1964 |
| | Floor space [$m^2$] | 5446.6 | 5447.41 | 4526 |
| Building energy performance indicators | Index of annual useful energy demand EU [kWh/(m2 · year)] | 103.63 | 110.00 | 67.17 |
| | Index of annual final energy demand EK [kWh/(m2 · year)] | 202.08 | 196.26 | 141.78 |
| | Index of annual primary energy demand EP [kWh/(m2 · year)] | 231.09 | 233.01 | 185.27 |
| | Specific volume of CO2 emissions ECO2 [t CO2/(m2 · year)] | 0.07 | 0.07 | 0.05 |
| | Share of renewable energy sources in annual final energy demand Uoze [%] | 0 | 0 | 0 |
| Area breakdown | Residential [$m^2$] | 3850.221 | 3975.7 | 4170.8 |
| | Other [$m^2$] | 1596.38 | 1471.71 | 355.2 |
| | Number of storeys | 5 | 5 | 5 |
| | Building volume [m3] | 13234.04 | 12905.35 | 11192.17 |
| | Lighting system | LED lighting (80%), Glow lighting (20%) | LED lighting (80%), Glow lighting (20%) | LED lighting (100%) |
| | External wall | 38cm grid brick + 10cm thermal insulation **U=0,31** | 38cm grid brick + 10cm thermal insulation **U=0,31** | 38cm grid brick + 10cm thermal insulation **U=0,31** |
| | Ceiling | Ceiling DZ3 + kermesite + 20cm mineral wool **U=0,17** | Ceiling DZ3 + kermesite + 20cm mineral wool **U=0,17** | Ceiling DZ3 + kermesite + 20cm mineral wool **U=0,17** |
| | Floor on the ground | Uninsulated floor on the ground **U=0,30** | Uninsulated floor on the ground **U=0,30** | Uninsulated floor on the ground **U=0,30** |
| | External window and balcony door | External double glazed window, insulated glass **U=1,30** | External double glazed window, insulated glass **U=1,30** | External double glazed window, insulated glass **U=1,30** |
| | External door | Aluminum external door, glazed **U=1,80** | Aluminum external door, glazed **U=1,80** | Aluminum external door, glazed **U=1,80** |

The figure presented below illustrates a technical drawing of a representative floor plan. The layout is characteristic of most floors, which tend to exhibit similar structural features.

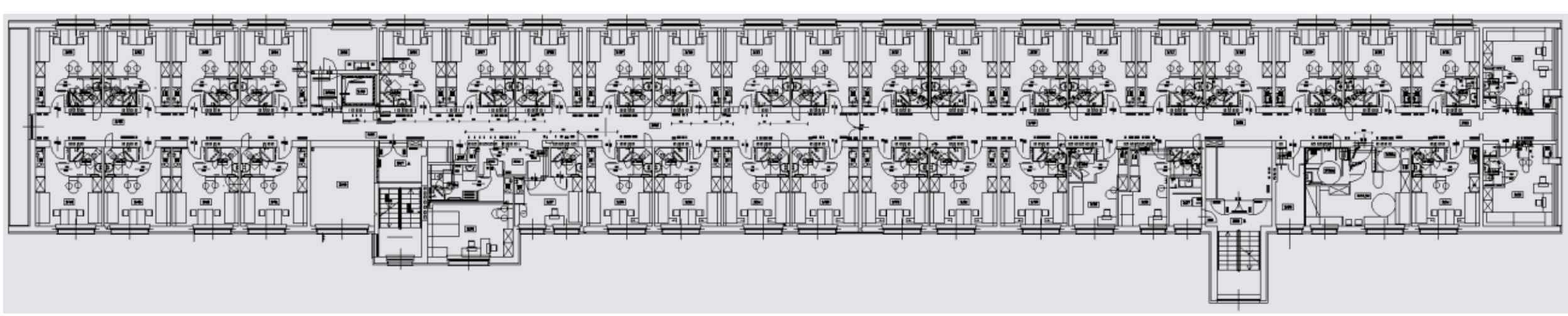

Floor plan of a sample floor in dormitory B.

**SM2:** Example mail sent to students: Framing participation as Collective Action

*Dear students,*

*This is a reminder that we are currently implementing further optimizations in our heating system aimed at reducing our P university carbon footprint. This initiative is a collective effort to explore various interventions that promise a significant reduction in CO2 emissions.*

*Why should you care? Your participation in this trial directly contributes to a sustainable future for us all. By joining forces, we can make a tangible impact on our environment while enjoying well-maintained living space in P university. During the last weeks and in the upcoming weeks, different heating interventions were and will be tested.*
*Remember, this is a joint venture! By realizing these actions, we pave the way for a greener, brighter future in P university.* ***Together, we are saving Co2.***